\documentclass[]{interact}
\usepackage{amsmath,amssymb, nccmath}
\usepackage{cuted}
\usepackage{algorithmic}
\usepackage{graphicx}
\usepackage{enumitem}
\usepackage[backend=bibtex,giveninits=true,citestyle=ieee]{biblatex} 
\usepackage{booktabs}
\usepackage{textcomp}
\usepackage{hyperref}
\usepackage{cleveref}
\crefformat{footnote}{#2\footnotemark[#1]#3}
\usepackage{siunitx} % https://tex.stackexchange.com/questions/269841/1-5e-10-style-scientific-notation-looks-ugly-in-latex-math-mode-how-can-i-forma
\usepackage{caption}
\usepackage[acronym]{glossaries}
\makeglossaries
\newacronym{isp}{ISP}{internet service provider}
\newacronym{poi}{POI}{point of interest}
\newacronym{od}{OD}{origin destination}
\newacronym{rog}{ROG}{radius of gyration}
\newacronym{gsm}{GSM}{Global System for Mobile Communication}
\newacronym{gps}{GPS}{Global Positioning System}
\newacronym{mno}{MNO}{Mobile Network Operator}
\newacronym{sdk}{SDK}{Sofware Development Kit}
\newacronym{cdr}{CDR}{Call Data Record}
\newacronym{xdr}{XDR}{X Data Record}
\newacronym{volte}{VoLTE}{Voice over LTE}
\newacronym{mo}{MO}{mobile originated}
\newacronym{mt}{MT}{mobile terminated}
\newacronym{clr}{CLR}{centered log-ratio}
\newacronym{pca}{PCA}{principal component analysis}
\newacronym{imsi}{IMSI}{international mobile subscriber identifier}
\usepackage{xcolor}
\renewcommand*{\bibfont}{\tiny}
\usepackage{todonotes}
\newif\ifdraftmode
\ifdraftmode
    \usepackage{draftwatermark}
\fi
    
\AtEveryBibitem{%
    {\clearfield{archivePrefix}
    \clearfield{arxivId}
    }
    {}%
}
\DeclareSourcemap{
  \maps{
    \map{
      \step[fieldset=eprint, null]
      \step[fieldset=eprinttype, null]
      \step[fieldset=arxivId, null]
      \step[fieldset=archivePrefix, null]
      \step[fieldset=issn, null]
      \step[fieldset=isbn, null]
    }
    \map{
      \pernottype{misc} % only if type is not misc
      \step[fieldset=url, null] % delete field
    }
  }
}

\begin{document}

\title{The impact of COVID-19 on relative changes in aggregated
mobility using mobile-phone data }

%%%%%%%%%%%%%%%%%%%%%%%%%%%%%%%%%%%%%%%%%%%%
% TODO add WWTF/CSH grant here
%%%%%%%%%%%%%%%%%%%%%%%%%%%%%%%%%%%%%%%%%%%%
%\thanks{
%Manuscript received December 1, 2012; revised August 26, 2015.
%This work was funded by the WWTF under project COV COV20-035.
%Corresponding author: P. Filzmoser (e-mail: peter.filzmoser@tuwien.ac.at).
%}
%}

%\author{\IEEEauthorblockN{anonymous authors}}

%%%%%%%%%%%%%%%%%%%%%%%%%%%%%%%%%%%
% simply copy  pasted authors. We still need to figure out who is an author
%%%%%%%%%%%%%%%%%%%%%%%%%%%%%%%%%%%

\author{
\name{Georg Heiler\textsuperscript{a,b}, Allan Hanbury\textsuperscript{a,b} and Peter Filzmoser\textsuperscript{c}\thanks{CONTACT Peter Filzmoser\textsuperscript{c}.
E-mail: peter.filzmoser@tuwien.ac.at}}
\affil{\textsuperscript{a}Institute of Information Systems Engineering, TU Wien, Favoritenstr. 9-11, 1040 Vienna, Austria; \textsuperscript{b}Complexity Science Hub, Josefstädter Str. 39, 1080 Vienna, Austria;
\textsuperscript{c} Computational Statistics
Institute of Statistics and Mathematical Methods in Economics, TU Wien, Wiedner Hauptstrasse 8-10
1040 Vienna
}
}

%This work was funded by the WWTF under project COV COV20-035.
%Corresponding author: P. Filzmoser (e-mail: peter.filzmoser@tuwien.ac.at).

%}

%\author{\IEEEauthorblockN{Georg Heiler\IEEEauthorrefmark{1}\IEEEauthorrefmark{2}, %2
%Allan Hanbury\IEEEauthorrefmark{1},
%Peter Filzmoser\IEEEauthorrefmark{1}
%,~\IEEEmembership{Fellow,~IEEE}
%}
%\IEEEauthorblockA{\IEEEauthorrefmark{1}Institute of Information Systems Engineering, TU Wien, Favoritenstr. 9-11, 1040 Vienna, Austria}
%\IEEEauthorblockA{\IEEEauthorrefmark{2}Complexity Science Hub, Josefstädter Str. 39, 1080 Vienna, Austria}
%}

\maketitle

\begin{abstract}
Evaluating relative changes leads to additional insights that would remain hidden when only evaluating absolute changes.
We analyze a dataset describing the mobility of mobile phones in Austria before, during COVID-19 lock-down measures until recently.

By applying compositional data analysis we show that formerly hidden information becomes available: we see that the elderly population groups increase relative mobility and that the younger groups, especially on weekends, also do not decrease their mobility as much as the others.
%\ifdraftmode
%    {\color{red} TODO write abstract}
%\fi
\end{abstract}


\section{Introduction}
\label{sec:intro}
The usage data of mobile-phones is used in a variety of different areas,
%to get deeper insight into the behavior of individual mobility,
such as during the COVID-19 pandemic \cite{Pepe2020, JaysonS2020, Gao2020, Jeffrey2020, Yabe2020,Vollmer2020,Xu2020, JRC2020b, JRC2020, JRC2020a, Heuzroth}, customer segmentation \cite{Aheleroff2011}, identification of personality traits and lifestyle \cite{Chittaranjan2011, Hillebrand2020}, the analysis of large social networks \cite{Aksu2019, Al-Molhem2019, Aledavood2018}, hotspot detection \cite{Nika2016}, prediction of movement \cite{Dao2019}, mode of transport identification \cite{Zhao2020}, credit scoring \cite{Liu2018}, disaster recovery \cite{Andrade, Marzuoli}, analysis of sleeping behavior of the population \cite{Monsivais2017a}, migration \cite{Isaacman2018} and land usage classification \cite{Chen2019a, Lenormand2015}.

The location of a mobile-phone is known for the \gls{mno} of the \gls{gsm} network.
We have partnered with an \gls{mno} in Austria to get access to such anonymized data.
%when this location information is stored over time.
We have defined an aggregation method to understand the overall aggregated mobility of the whole population.
Our data set as well as the aggregation, anonymization approach and the various phases of the lock-down are outlined in detail in \cite{heiler2020covidbasic}.

%WeThe interest here is not on an individual level, but in aggregated information, for example 
%aggregated for gender or different age groups. Most importantly, the aggregation
%level needs to be taken such that data privacy can be guaranteed.
%UND NOCH MEHR DAZU ...

With the outbreak of COVID-19 and the subsequent lock-down in Austria, the mobility
behavior of the population has changed significantly. 
This is reflected in the mobility data derived from mobile-phone information \cite{heiler2020covidbasic}.
To measure mobility, an appropriate measure needs to be established which reflects the mobility.
One possibility is the \gls{rog} \cite{Gooch2011}.
It is formally defined below, and refers to the time-weighted distance of the movement locations to the main location.
We compute it on a daily level.
Its unit is meters, and the values
are strictly positive.
In this work the analyze the aggregated (median) \gls{rog} of the whole population of Austria for various groups as a time series.
The groups are defined by gender- or age groups.

Traditionally, a comparison is made in terms
of \textit{absolute information}, i.e., the \gls{rog} time series 
values of the different groups are analyzed in their unit of meters.
We have conducted such an analysis \cite{heilerAndReish2020CovidGender} which focuses on gender differences.

An alternative is to compare \textit{relative information}, for example the
\gls{rog} of the males with respect to females, or in terms
of the ratio males to females.
This leads to a dimensionless time series, and to a different aspect of data analysis which emphasizes the differences between the individual groups. 
A joint increase or decrease in both groups may not lead to a big change of the ratio.
On the other hand, the ratio will change if the values of one group increase, and at the same time they decrease in the other group, or vice versa.
Here again, the relative change rather than the absolute change is important.
For example, if the \gls{rog} changes from 1000m to 2000m in one group, and
from 2000m to 1000m in the other group, the ratio would change from 1/2 to 2.
The same change could be observed if the absolute values in both groups would be bigger by a factor of 10.
Thus, absolute values are no longer
relevant in this consideration, because a multiplication by any positive 
constant leads to the same ratio.
This is still trivial in case of comparing two groups, but it is no longer
straightforward when relative information of several groups, such as age classes,
should be compared.
\emph{Compositional data analysis} is devoted to this problem
of analyzing relative information \cite{Aitchison86,pawlowsky15,filzmoser2018applied}. 
In fact, compositional data analysis is frequently used in geosciences, but also more and more in
other fields such as 
biology \cite{espinoza2020applications}, 
bioinformatics \cite{quinn2018understanding}, 
economics \cite{trinh2019relations}, 
marketing \cite{joueid2018marketing}, 
medicine \cite{dumuid2020compositional}, etc.     

We contribute compositional analysis of the movement data during the COVID-19 analysis which comes to the conclusion that special groups (elderly and young cohorts during weekends) need an additional caring treatment to improve their security. 

This work is structured as follows.
In Section~\ref{sec:coda} we give a 
brief mathematical introduction to compositional data
analysis.
Section~\ref{sec:data} provides more details about the mobile-phone data used and about the quantities derived.
In addition to the mobility measured as the \gls{rog}
we will furthermore investigate the call duration per day, again aggregated by the median.
Section~\ref{sec:results} presents comparisons of the analysis based on absolute and on relative information, and the final Section~\ref{sec:conclusions} summarizes the findings.% and concludes.

%%%%%%%%%%%%%%%%%%%%%%%%%%%%%%%%%%%%%%%%%%%%%%%%%%%%%%%%%
\section{Materials and methods}
% https://www.tandfonline.com/action/authorSubmission?show=instructions&journalCode=cjas20
% adapted to suggestions of paper
\subsection{Compositional data analysis}
\label{sec:coda}
From the point of view of compositional data analysis, a composition is defined as multivariate information, consisting of strictly positive values, where the absolute numbers as such are not of interest, and only relative
information is relevant for the analysis
\cite{filzmoser2018applied}.
A composition can be given for example by the median \gls{rog} values of different age categories for a certain day, and every age category is denoted as a compositional part.
We use the notation $x_1, \ldots ,x_D$ for the compositional parts of $D$ categories,
and the composition is written as the (column) vector $\mathbf{x}=(x_1,\ldots ,x_D)'$. 
For every day recorded in our data base we will observe such a composition, which
in fact leads to a multivariate compositional time series.
The interest is in relative information in terms of the ratios, and thus all pairs $x_j/x_k$, for $j,k= 1,\ldots ,D$, should be considered in the analysis.
Obviously, the pairs for $j=k$ are not relevant, and pairs of the reverse ratio $x_k/x_j$ do not contain potentially new information.
This motivates to consider the logarithm of the ratios, $\ln(x_j/x_k)$, so-called log-ratios.
The reverse ratios have a different sign, and thus do not need to be considered, and their variance is the same as for the original ratio.
Moreover, the distributions of log-ratios tend to be more symmetric than without a logarithm \cite{pawlowsky15}.

Still, the resulting $D(D-1)$ pairs $\ln(x_j/x_k)$, for $k>j$, only live in a subspace of dimension $\leq D-1$ \cite{filzmoser2018applied}, and thus it is natural to aggregate this information.
Consider an aggregation
\begin{equation}
\label{eq:clr1}
y_1=\frac{1}{D}\left(\ln\frac{x_1}{x_2}+ \ldots + \ln \frac{x_1}{x_D}\right)=
\ln\frac{x_1}{g(\mathbf{x})} \ ,
\end{equation}
where 
$$
g(\mathbf{x})=\sqrt[D]{\prod_{j=1}^D x_j}
$$ 
is the geometric mean of the composition $\mathbf{x}$.
Then, $y_1$ represents all relative information about the part $x_1$ to the other 
parts in the composition in a form of an average of the log-ratios. 
This leads to the definition of so-called \gls{clr} coefficients \cite{Aitchison86}
\begin{equation}
\label{eq:clr}
\mathbf{y}=(y_1,\ldots ,y_D)' 
\quad \mbox{ with } \quad
y_j=\ln \frac{x_j}{g(\mathbf{x})} \ .
\end{equation}
The vector $\mathbf{y}$ contains all relative information about $\mathbf{x}$ in the above sense.
It consists of $D$ components $y_j$ which are associated with the relative information about the corresponding part $x_j$.
However, it turns out that $y_1+\ldots +y_D=0$, and thus a representation of data in terms of \gls{clr} coefficients leads to singularity \cite{filzmoser2018applied}.
Although there are ways to circumvent this issue \cite{filzmoser2018applied}, we will proceed with \gls{clr} coefficients for the following analysis for simplicity.

Consider now a multivariate compositional time series
$\mathbf{x}_{t}=(x_{t1},\ldots ,x_{tD})'$, for the time points $t=1,\ldots ,T$,
and the observations $x_{tj}$ for each part $j\in \{1,\ldots ,D\}$.
The time series expressed in \gls{clr} coefficients is
$\mathbf{y}_t=(y_{t1},\ldots ,y_{tD})'$, with $y_{tj}=\ln(x_{tj}/g(\mathbf{x}_t))$,
with the geometric mean $g(\mathbf{x}_t)=(\prod_{j=1}^D x_{tj})^{1/D}$ per time point.
Since this data representation only reflects relative information of the time series, an additional visualization of the absolute time series values can be interesting
to get a more complete picture.

The \gls{clr} coefficients result in multivariate data that can be analyzed with the traditional multivariate statistical methods \cite{filzmoser2018applied}.
A prominent way to represent the information in a lower-dimensional space is
to use \gls{pca}.
Since PCA is sensitive to data outliers or inhomogeneous data, robust versions have been proposed, also in the compositional data analysis framework \cite{filzmoser2018applied}.
The resulting loadings and scores are commonly represented in a biplot
    to get an overview of the multivariate data \cite{AG02}.

%%%%%%%%%%%%%%%%%%%%%%%%%%%%%%%%%%%%%%%%%%%%%%%%%
\subsection{Mobile-phone data}
\label{sec:data}

In this work we analyze two measures obtained from the mobile phone data, the call duration and the radius of gyration \gls{rog}.
While the meaning of the former is straightforward, the latter needs
to be defined.

Consider an individual $i:=i(t)$ at a certain day 
$t\in\{1,\ldots ,T\}$.
For reasons of data privacy, the index of the individual will change every day.
The  data is made available to the researchers already anonymized with a daily changing key.

Furthermore, the current location of the individual's mobile phone is available at the time points $t_{\tau}=t+\tau_t$, for a number of  time points $T_t$ per day, where $\tau_t \in[0,1)$.
The corresponding $x$- and $y$-coordinates are denoted by $(\xi_{it_{\tau}}, \eta_{it_{\tau}})$.
With this information, the stay duration $l_{it_{\tau}}$ for individual $i$ at time point $t_{\tau}$ can be computed, which is used to calculate a weighted average
$(\bar{\xi}_{it},\bar{\eta}_{it})=
\left(
\frac{\sum_{t_{\tau}}l_{it_{\tau}}\xi_{it_{\tau}}}
{\sum_{t_{\tau}}l_{it_{\tau}}},
\frac{\sum_{t_{\tau}}l_{it_{\tau}}\eta_{it_{\tau}}}
{\sum_{t_{\tau}}l_{it_{\tau}}}
\right)$
for individual $i$ for day $t$.
These coordinates are in the middle of the area covered by  all the locations which were visited during the day and are dominated by the two most prominently used (longest used) locations: home and work location.
%These coordinates approximate the most used location (if the time points are taken during the evening and night hours this most likely constitutes a home location), or the location of the working place (if the time points are taken during the day). 

Denote $d^2_{it_{\tau}}=(\bar{\xi}_{it}-\xi_{it_{\tau}})^2
+(\bar{\eta}_{it}-\eta_{it_{\tau}})^2$ as the squared Euclidean distance
between the coordinates $(\bar{\xi}_{it},\bar{\eta}_{it})$ and 
$(\xi_{it_{\tau}}, \eta_{it_{\tau}})$.
This requires the coordinate system to be local in order to obtain valid, i.e. less distorted results.
Otherwise, a Haversine distance could be used instead in case of epsg:4326 WGS-84 projection of the coordinates.

The \gls{rog} for individual $i$
and day $t$ is then defined as 
\begin{equation}
R_{it} = \sqrt{\frac{\sum_{t_{\tau}} d^2_{it_{\tau}}}
{\sum_{t_{\tau}} l_{it_{\tau}}}} ,
\label{eq:rog}
\end{equation}
and it thus represents a distance to the center of all the places of stay during that day $t$ weighted by the lengths of the stay duration at the different places.

Details are available and especially a description of how the large quantity of data was handled is available in \cite{heiler2020covidbasic}.
%\ifdraftmode
%    {\color{red} cite gender paper. Should we talk about and improve it with \gls{clr}?}
%\fi

Using additional metadata, an individual $i$ can be assigned to a gender
group (female, male), to an age group (here we consider the age groups in 15 year intervals:
15-29, 30-44, 45-59, 60-74, and 75$+$), and to an Austrian district of the daily night location to derive the groups.
%With this information, it is possible to summarize the distribution of the \gls{rog} or of the call duration per considered grouping variable(s) and day.
Since the distributions are generally very right-skewed, we work with the median per group and day in the following and also ensure k-anonymity for each one.
%The \gls{rog} of the anonymized individuals are aggregated by calculating the median and ensuring k-anonymity for each cohort.

The resulting time series can be directly investigated in terms of  their absolute information, and they can be compared to an analysis based on relative information.

%%%%%%%%%%%%%%%%%%%%%%%%%%%%%%%%%%%%%%%%%%%%%%%%%%%%%%%%%%%%%%%%%%%%%%%%%%%%%%%%%
\section{Results}
\label{sec:results}

\subsection{Mobility measured by \gls{rog}}
\begin{figure}%[htbp]
\centerline{\includegraphics[width=\linewidth]{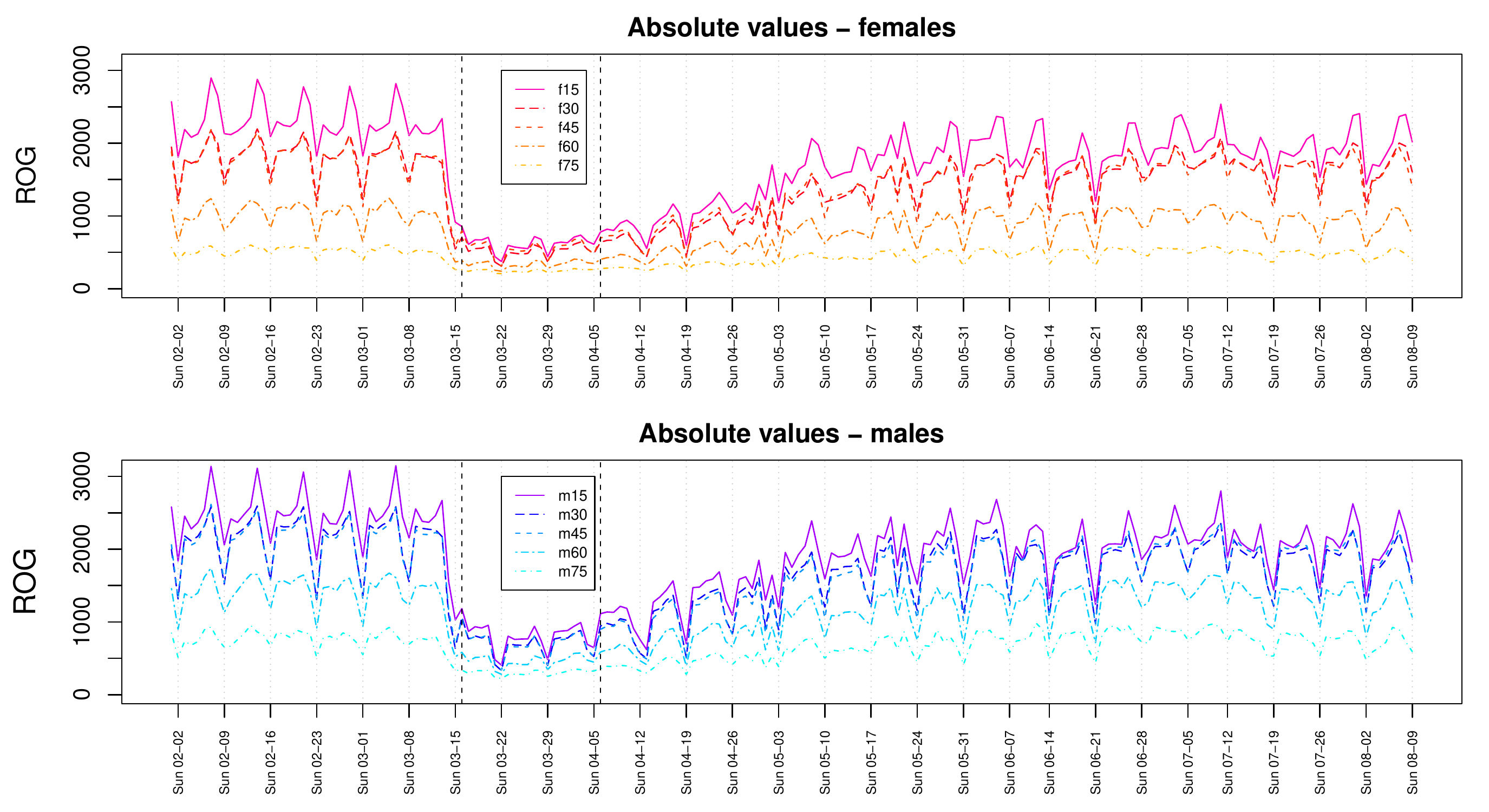}}
\caption{Median \gls{rog} values for different age groups over time
for females (top) and males (bottom) in different age groups.}
\label{fig:plot_r50abs}
\end{figure}

The results reported in this section refer to the median values of
the \gls{rog} per group.
To begin with, Figure~\ref{fig:plot_r50abs} shows the absolute values for the females (top) and males (bottom) for different age groups.
The legend indicates the considered age groups: 15 for age 15-29, 30 for age 30-44, 45 for age 45-59, 60 for age 60-74, and 75 for age elder than 75. For all of the following time series  plots, the vertical dashed lines indicate the date March 16\textsuperscript{th}, 2020, when the restrictions came into action, and the date April 6\textsuperscript{th}, 2020, when they were relaxed. 
The data considered here are from the period February 1\textsuperscript{st} until August 9\textsuperscript{th}, 2020.
The lock-down is clearly visible in the plots by an abrupt decay of the median \gls{rog} values in all age classes for both genders. After the lock-down, the order of the values still remains the same, from the eldest group with the smallest values, and the youngest group with the highest values, but it is on a much smaller level. The level then increased more or less systematically until the middle of June. Afterwards, the level
is not changing a lot, it is lower than at the beginning, and weekly patterns are clearly visible.
Note that these weekly time series patterns that are very regular at the beginning are getting somehow distorted, partially also due to holidays (April 13\textsuperscript{th}, May 1\textsuperscript{st}, May 21\textsuperscript{th}, June 1\textsuperscript{st}, June 11\textsuperscript{th}), and they never get back to this regularity.

\begin{figure}%[htbp]
\centerline{\includegraphics[width=\linewidth]{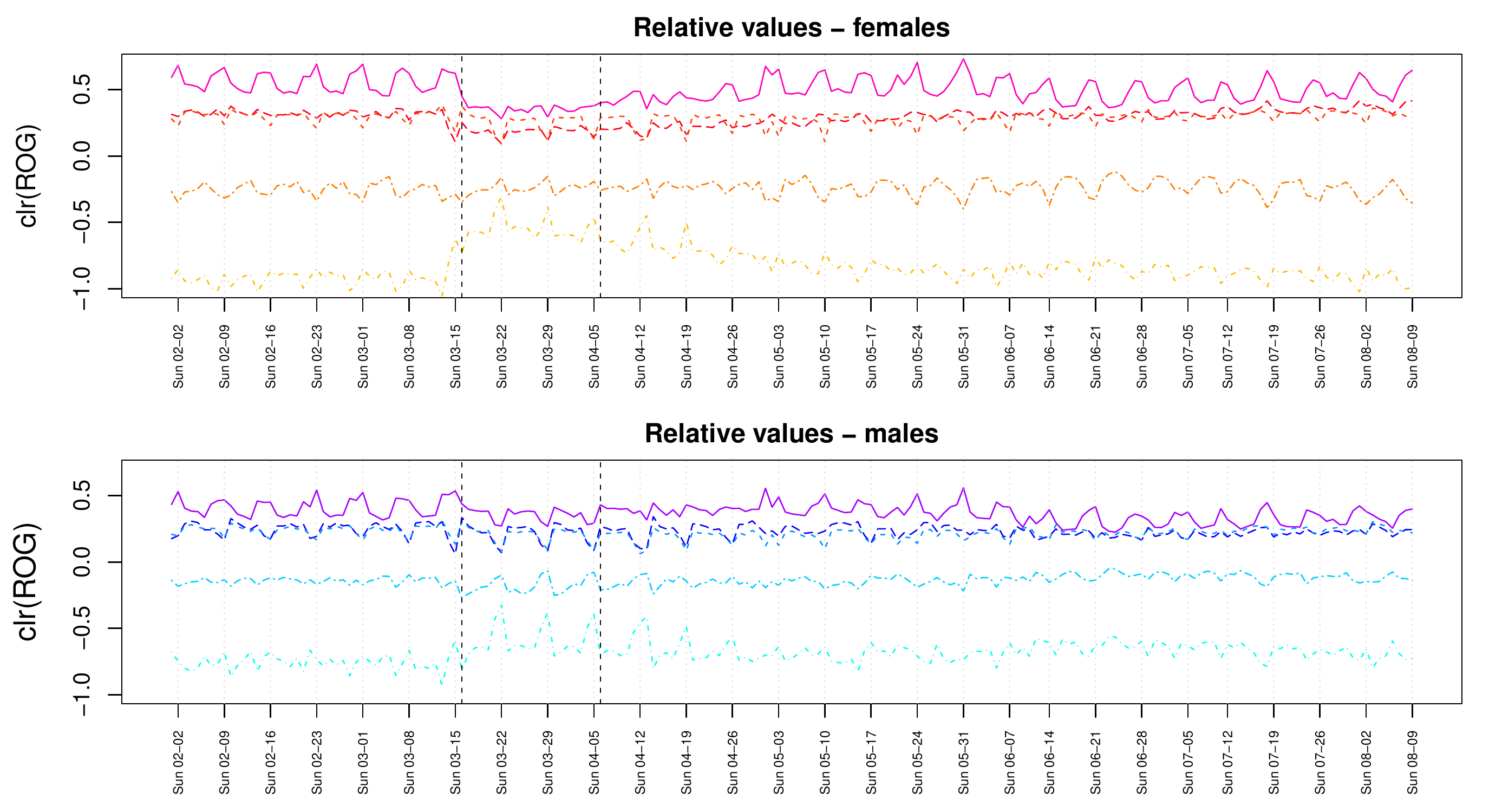}}
\caption{\gls{clr} coefficients of median \gls{rog} values for the female (top) and the male (bottom) composition.
For the legend see Figure~\ref{fig:plot_r50abs}.}
\label{fig:plot_r50rel}
\end{figure}

Figure~\ref{fig:plot_r50rel} focuses on the relative information 
contained in the median \gls{rog} values.
We consider the female age groups and the male age groups separately as two compositions.
The plots show the corresponding \gls{clr} coefficients for females (top) and males (bottom). While in Figure~\ref{fig:plot_r50abs} we have essentially
seen a decline of all values at the beginning of the lock-down phase,
followed by an increase, we did not pay attention how differently
the age groups declined and increased. This is the purpose of 
the relative view in Figure~~\ref{fig:plot_r50rel}, where we mainly 
investigate the developments of the age groups to each other.

In both plots of Figure~~\ref{fig:plot_r50rel} we can see roughly the same pattern after the lock-down: the biggest relative changes are visible for the youngest and the oldest age group, but they go into different directions.
While group 15 had the biggest decline, group 75$+$ increased the values relative to the other age groups.
This seems to be counter-intuitive, but it can be explained by the fact that the geometric mean also went down significantly, and the ratio of the values of group 75$+$ to the geometric mean then even increased after the lock-down.
Another interesting phenomenon is that the groups 60 and 75$+$ show the biggest increase in mobility (in a relative sense) during the weekends in this lock-down period. Although on a different level, the values from July
show a similar structure to those from February. It is interesting to note 
that the youngest age group 15 shows a somehow mirrored weekly pattern
compared to the elder age groups. This is not visible when looking at the
absolute values in Figure~\ref{fig:plot_r50abs}.

\begin{figure}[htbp]
\centerline{\includegraphics[width=\linewidth]{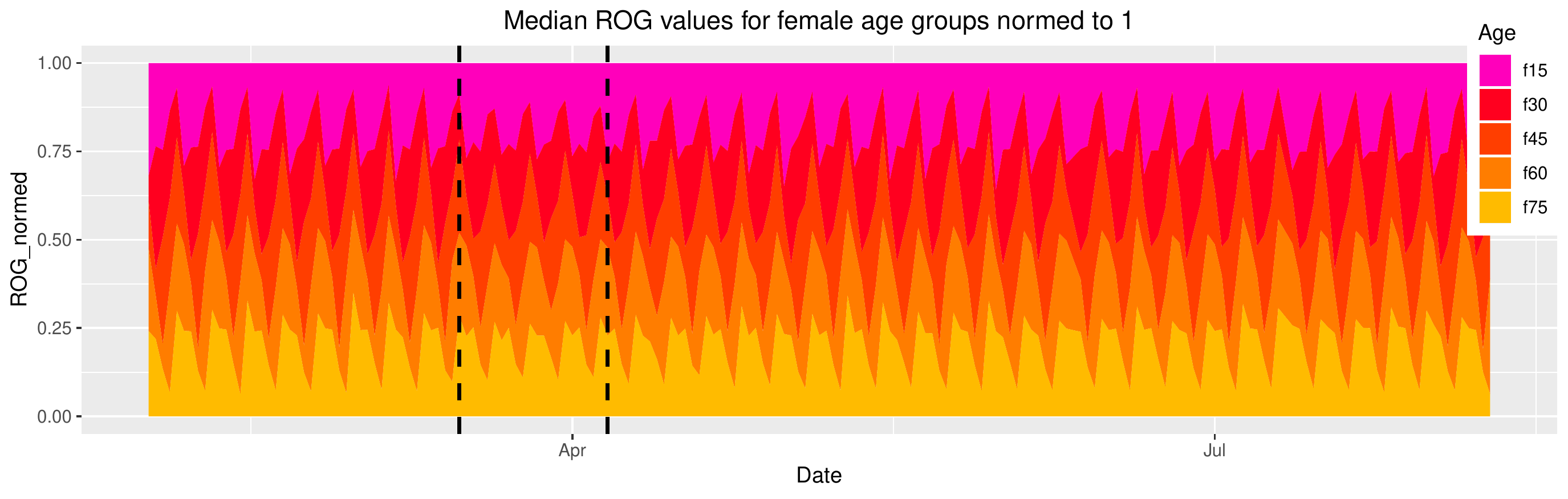}}
\caption{Proportional presentation of the median \gls{rog} values for the 
female age groups. For each time point, the data are normalized to a value 
of 1.}
\label{fig:plot_r50prop}
\end{figure}

Relative information could also be understood in terms of data proportions.
In particular, one could compute the proportion of a group on the
total per time point, which in fact corresponds to normalizing the
data per time point to a value of 1.
Such a proportional presentation
is shown in Figure~\ref{fig:plot_r50prop} for the \gls{rog}
values of the female age groups.
Obviously, the information contained in
this representation is different from \gls{clr} coefficients which focus on
log-ratio information.
One can hardly see any differences between the
lock-down period and the remaining period, and thus this kind of
``relative view'' is not valuable for the analysis.

\begin{figure}%[htbp]
\centerline{\includegraphics[width=\linewidth]{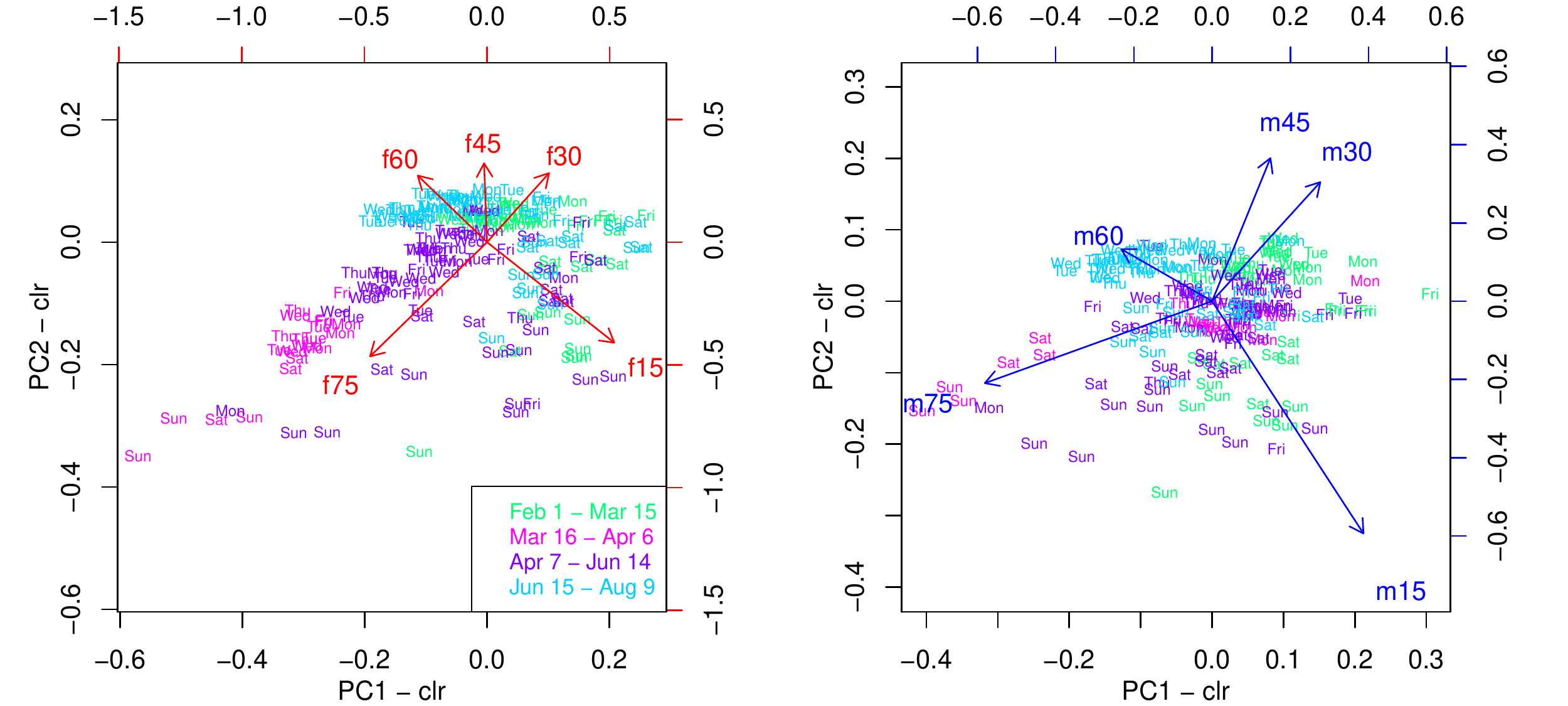}}
\caption{Biplots of the \gls{clr} coefficients of the median \gls{rog} values 
for females (left) and male (right) age groups. Green color for period before the lock-down, pink for lock-down period, purple after lock-down until mid of June, and light-blue after this period.}
    \label{fig:biplot_r50}
\end{figure}

The median \gls{rog} values for the female and male age groups are analyzed
in the following with PCA. Here, the method ROBPCA \cite{hubert-ROBPCA:2005}
is taken, a robust version of PCA which downweights outlying observations.
Figure~\ref{fig:biplot_r50} shows the biplot of the first two principal components (PCs)
for the clr coefficients. The coloring is according to the time phases:
green before the lock-down, pink during the lock-down period, purple
after lock-down until mid of June, and 
light-blue after this period. The left biplot for the females
identifies these four periods as clear clusters, while there is more
overlap visible in the right biplot for the males. 
For the females, the direction of the first PC (71\% explained
variance) shows a transition
of the relative \gls{rog} values from the young generation (f15, f30) 
before lock-down to the old (f75) one during lock-down, and then back to the 
center. Thus, younger and elder females show a contrasting behavior in this
time period, which was already observed in Figure~~\ref{fig:plot_r50rel} (top panel).
The second PC (21\% explained variance) shows also differences between
the time periods, but it also reveals weekend effects. Especially
on Sundays, the mobility for group f15 was
bigger before and after lock-down, but it moved to group f75 during the
lock-down phase. 

The data structure in the biplot for the males (right plot) looks a bit
different, but leads to similar conclusions. PC1
explains 69\% and PC2
25\% of the variance. Groups m75 and m15 have a similarly diverging 
behavior of Sunday mobility as observed for the females.
The weekdays of the lock-down phase are in the center of the distribution,
while for the females they were clearly moved towards group f75.
On the other hand, the weekdays in the first time period (February 1\textsuperscript{st} - March 15\textsuperscript{th})
are better distinguishable from the weekdays of the last period (June 15 -
August 9); a possible explanation is the fact that the working male
population changed the mobility behavior more significantly than
that of females due to home office.

\begin{figure}%[htbp]
\centerline{\includegraphics[width=\linewidth]{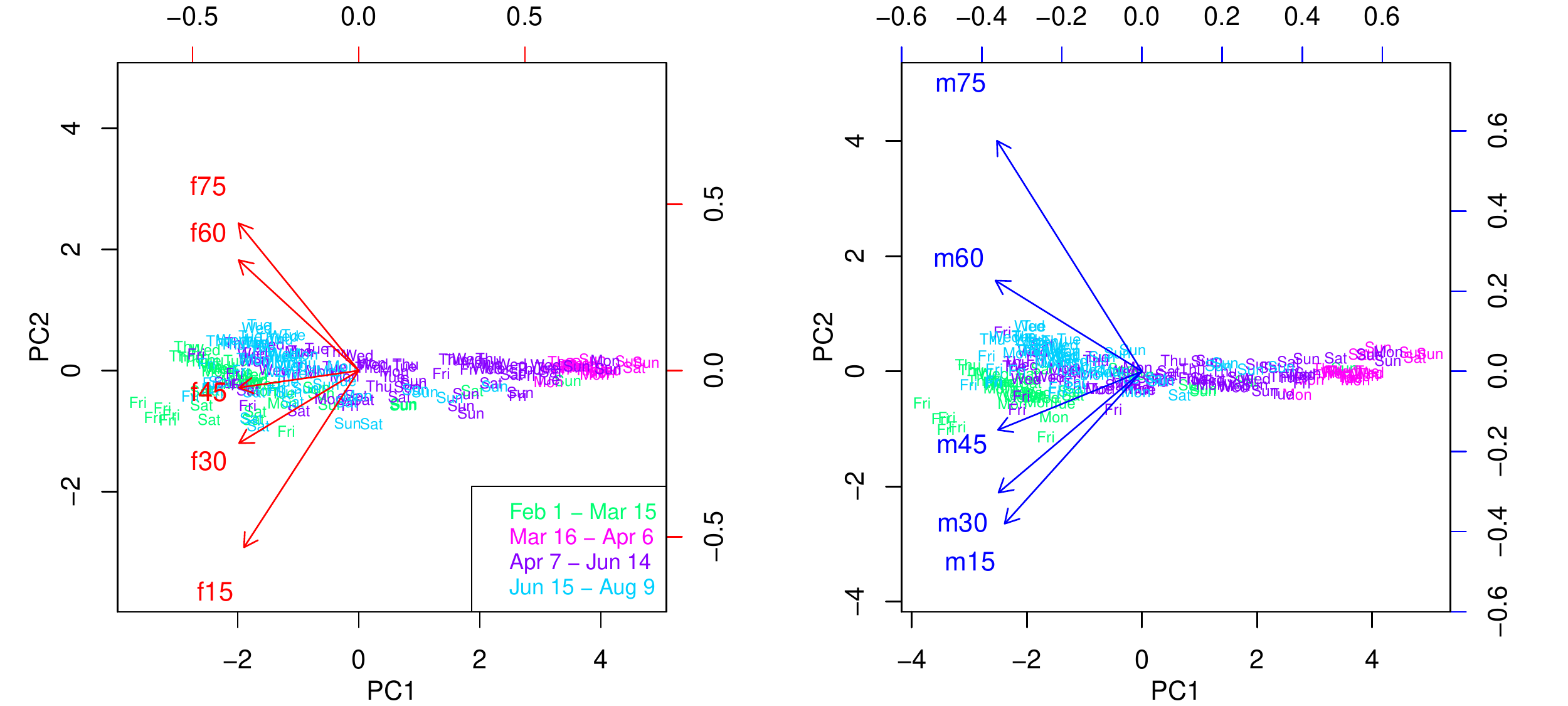}}
\caption{Biplots of the (absolute) median \gls{rog} values 
for females (left) and male (right) age groups. Green color for period before the lock-down, pink for lock-down period, purple after lock-down until mid of June, and light-blue after this period.
}
    \label{fig:biplot_r50abs}
\end{figure}

A quite contrasting view is revealed in Figure~\ref{fig:biplot_r50abs}, which
shows the robust PCA results for the absolute
values of \gls{rog}, for females (left) and males (right). In both
analyses, PC1 explains 98\% of the variability, and this direction 
essentially reflects the big change of the \gls{rog} over this time 
period. Otherwise, there is not much information left in these analyses,
which reflects the limited usefulness of absolute information if the task
is to compare age groups.

\subsection{Interaction measured by call duration}

\begin{figure}%[htbp]
\centerline{\includegraphics[width=\linewidth]{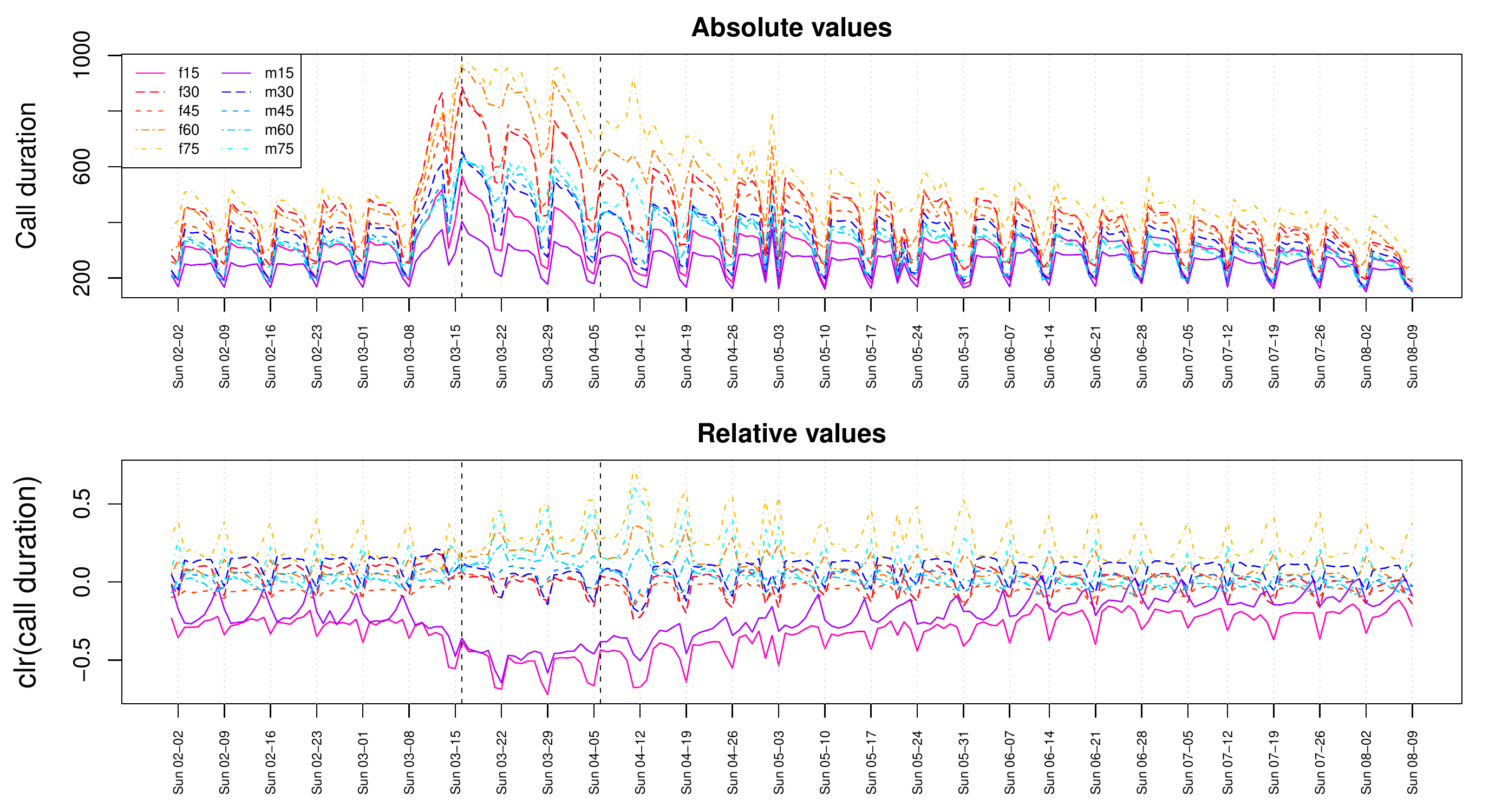}}
\caption{Median values of call durations per gender and age group over time (top),
and clr representations separately for female and male age groups (bottom).}
\label{fig:plot_mo50}
\end{figure}

Figure~\ref{fig:plot_mo50} investigates the median call duration, reported
in seconds, again for the two genders and the age groups. The absolute values
are shown in the upper plot jointly for males and females. Here we observe the 
reverse ordering of the age groups compared to the plots for the \gls{rog}
values: the lowest values are for the youngest group, and the biggest for the
oldest group. The values of the females are systematically higher than those
of the males. It is interesting to see that the call durations already started
to increase one week before the lock-down. While the \gls{rog} time series had
their peaks during the weekend, we have the opposite here. This pattern, however,
seems to change after the lock-down for group f75 (uppermost line),
and it went back to \emph{normality} only later on.

The bottom plot of Figure~\ref{fig:plot_mo50} presents the \gls{clr} coefficients, which
are separately calculated for females and males, but presented here jointly for
easy comparison. Although the absolute values of the youngest age group
also increased with the lock-down, the increase was smaller compared to the 
other groups, which is reflected by decreasing \gls{clr} coefficients. The pattern
of f15 and m15 has also an interesting structure: Before the lock-down,
the groups had quite different behavior within their gender-group, but during
the lock-down phase they became quite similar. From June on, they show again a 
similar behavior as at the beginning. Another interesting phenomenon can be 
seen after the lock-down: the two oldest groups show a contrary behavior to
the other groups during the weekends. Their decline in call duration during
the weekends was much smaller than that of the other age groups.

\begin{figure}%[htbp]
\centerline{\includegraphics[width=\linewidth]{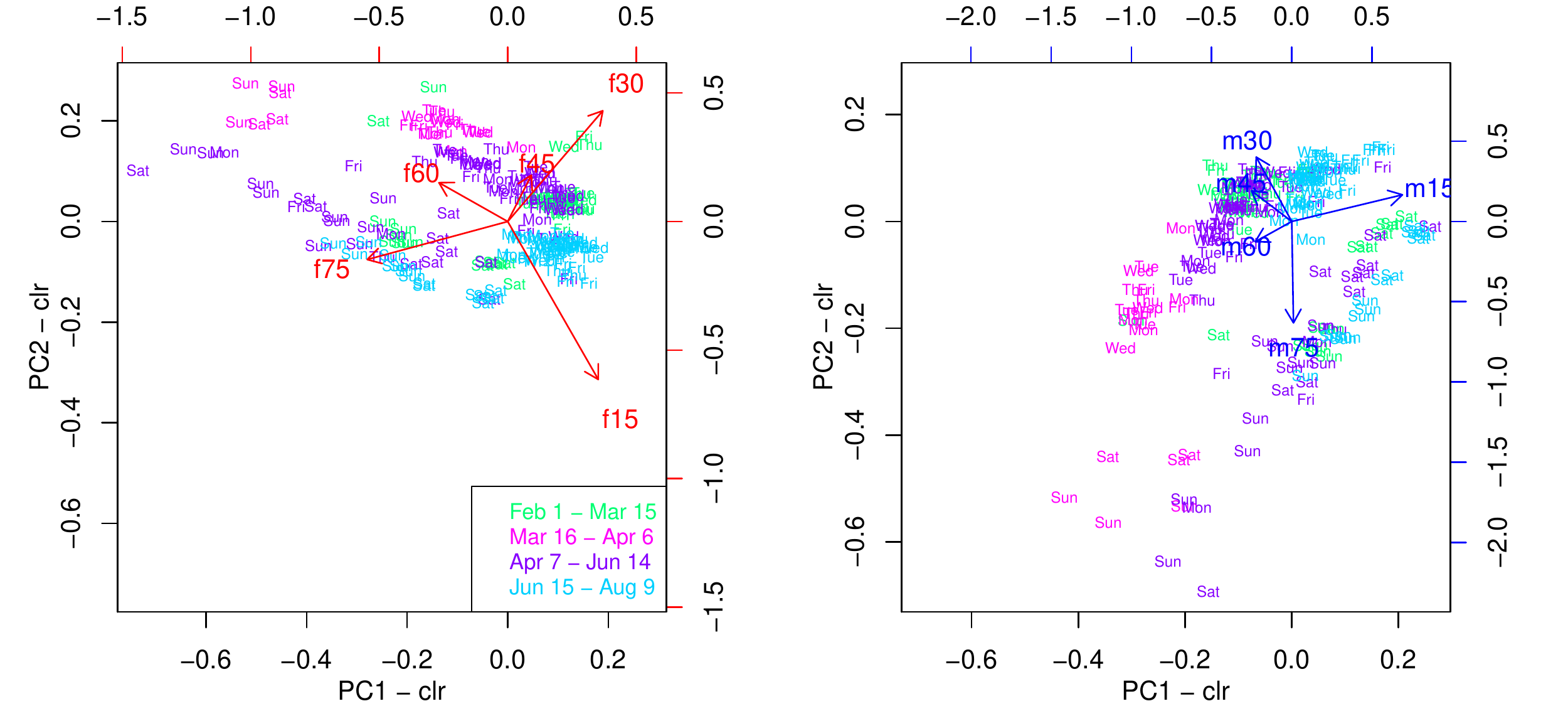}}
\caption{Biplots of the \gls{clr} coefficients of the median call duration values 
for females (left) and male (right) age groups. Green color for the period before the lock-down, pink for lock-down period, light-blue after this period.}
    \label{fig:biplot_mo50}
\end{figure}

Figure~\ref{fig:biplot_mo50} presents biplots of a robust PCA for the \gls{clr}
coefficients for the female (left) and male (right) age groups. The coloring
is taken as in the previous biplots, green before lock-down, pink during, 
purple after lock-down, and light-blue from June 15\textsuperscript{th} onwards. 
PC1 explains 72\% of the variability for the 
females, 54\% for males, and PC1 and PC2 together explain about 98\% variance
in both cases. The different groups which are visible in the biplots are
essentially weekend-effects or affects due to the lock-down. 
These grouping effects are essentially caused by the youngest and
oldest age groups.
When comparing the first observed time period with the last one,
we can find quite clear differences in the corresponding PCA scores.
These differences are essentially caused by the changing contrasting 
behavior between the youngest group and the elder groups; groups
f75 and m75 (and also m30) do not seem to contribute to this difference.
A possible explanation is the exploration of alternative methods for
communication, especially for the elder groups.

\subsection{Interactions between source and destination}

It can be recorded who is actively calling a person, and who is receiving a call.
The former person is called \emph{source}, and the latter \emph{destination}. 
Here we investigate the median \gls{rog} values for the different age groups of the females and males.
However, the data set is more complex now, because a person from a certain age group can be source, while the destination can originate from a different age group.
Moreover, both source and destination will have
specific median \gls{rog} values.

\begin{figure}[htbp]
\centerline{\includegraphics[width=\linewidth]{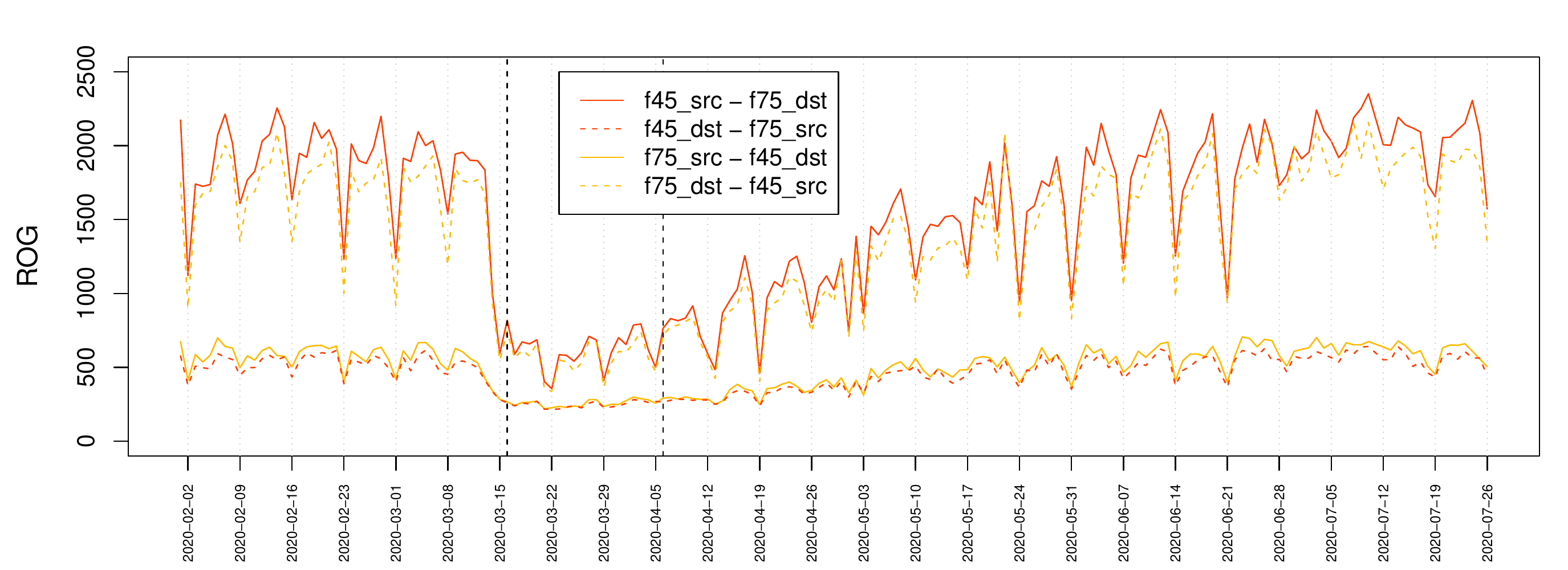}}
\caption{Median \gls{rog} values for the female age groups $f45$ and $f75$,
depending whether they actively call (src) or they passively receive
the call (dst). For example, line f75$\_$src -- f45$\_$dst refers to the 
median \gls{rog} values for females in age group f75, actively calling
females in age group f45.}
    \label{fig:plot_45_75}
\end{figure}

Figure~\ref{fig:plot_45_75} illustrates these data for four specific cases:
source f45 (f45$\_$src) with destination f75 (f75$\_$dst), and source 
f75 (f75$\_$src) with destination f45 (f45$\_$dst). In both cases, the 
median \gls{rog} values can be taken from the source group or from the 
destination group, see also
figure legend. Throughout the whole period (here from February 1\textsuperscript{st} - July 26\textsuperscript{th}), the median \gls{rog} values from 
the source groups (solid lines) have slightly higher
values than those of the destination groups (dashed lines) for the same 
age classes, which can be expected because people from the source groups might
call from a place outside their usual environment. While the lines 
are on a similar level at the beginning and at the end of the considered
period, the weekly periodicity changes, probably caused by the 
summer holidays.

\begin{figure}%[htbp]
\centerline{\includegraphics[width=0.48\linewidth]{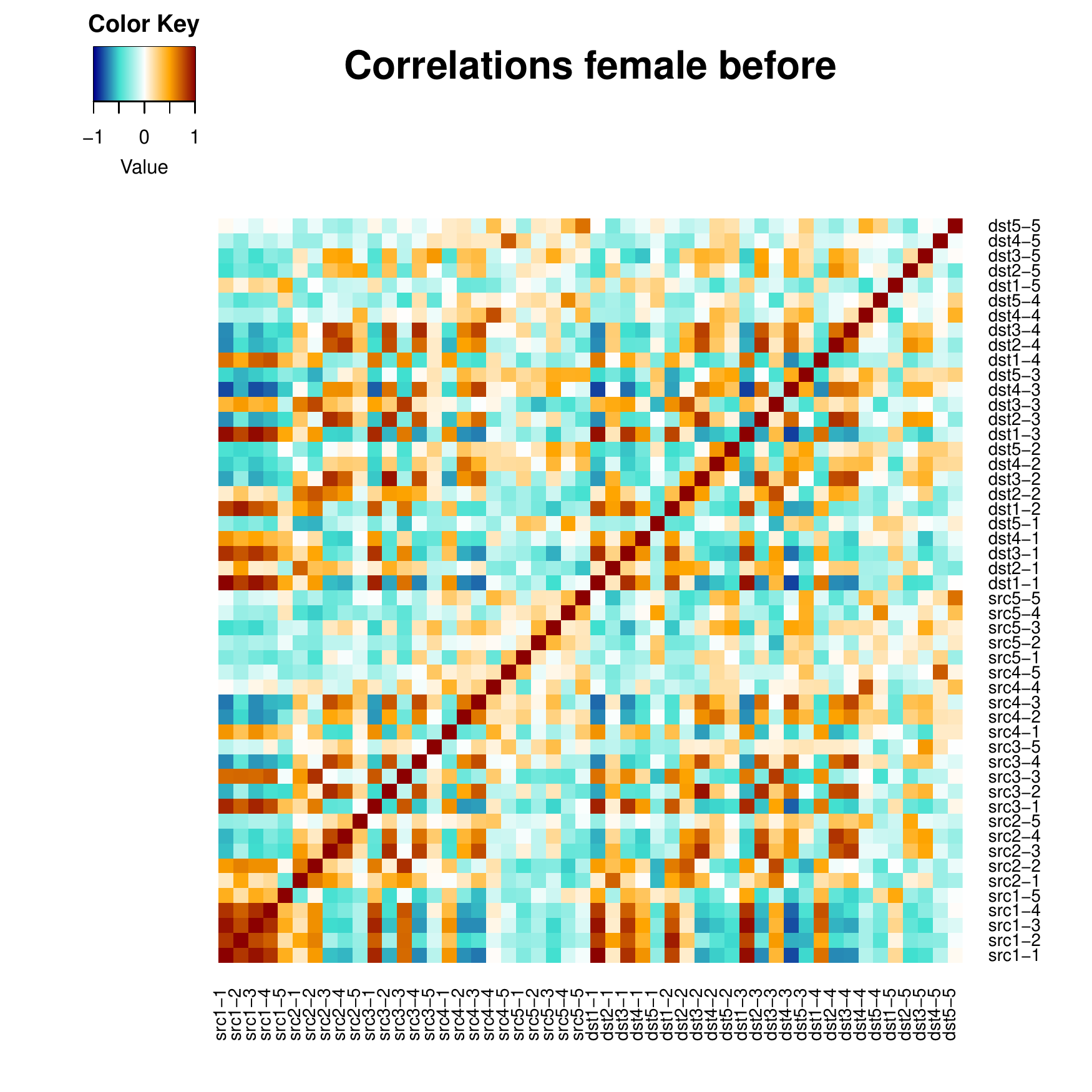}
\includegraphics[width=0.48\linewidth]{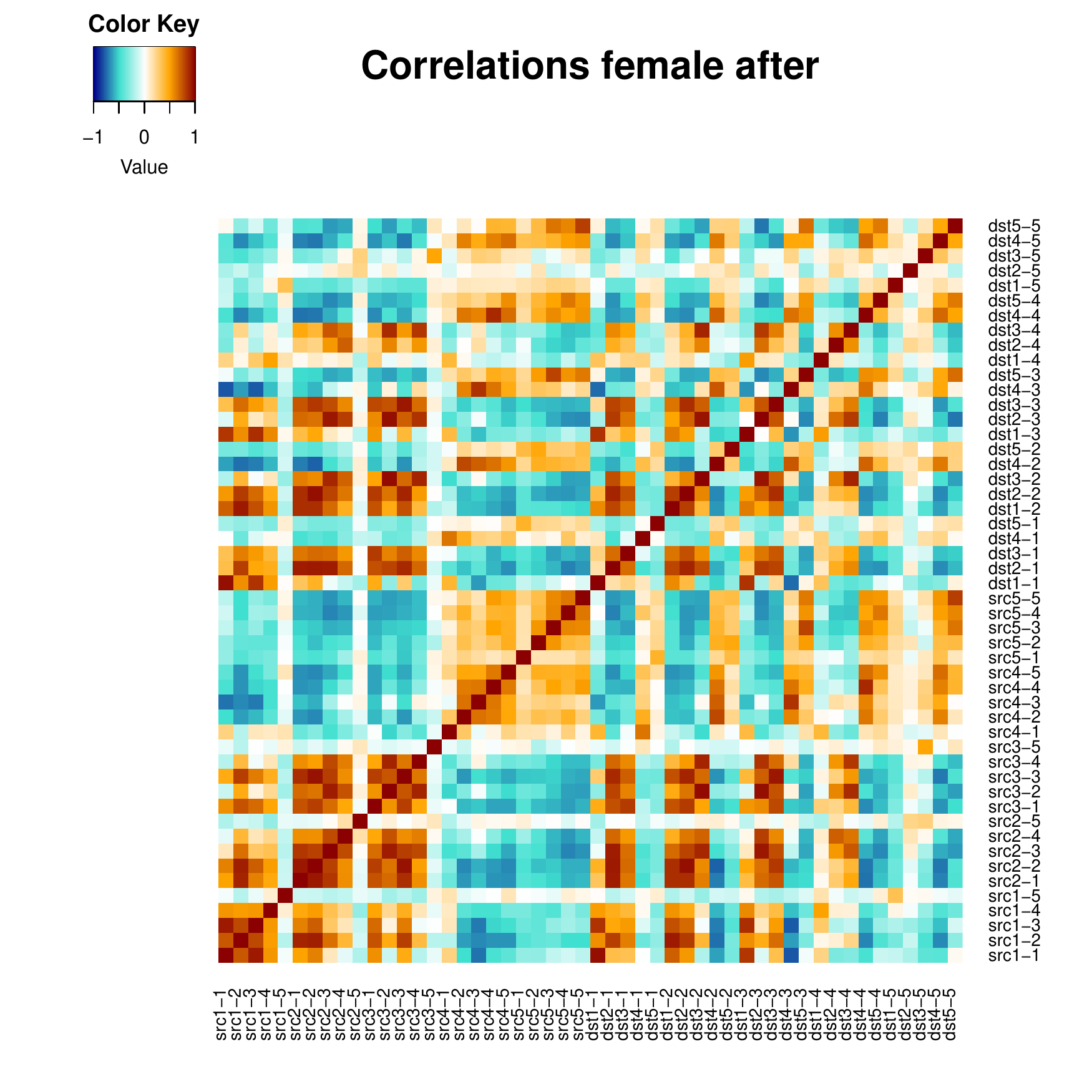}}
\caption{Correlations of the \gls{clr} coefficients for median \gls{rog} values 
for the female age groups (1 to 5, referring to 15 to 75$+$),
when they are actively calling (src) or passively receiving a call (dst), 
recorded before March 16\textsuperscript{th}, 2020 (left), and afterwards (right).}
    \label{fig:plotcorf}
\end{figure}

In the following analyses we are interested in the similarity of the relative
\gls{rog} values in terms of correlations, before
lock-down (February 1\textsuperscript{st} -- March 15\textsuperscript{th}) and after (March 16\textsuperscript{th} -- May 31\textsuperscript{th}). In order to 
investigate relative information, the \gls{clr} coefficients are computed for
a composition with all 25 age combinations of the source-destination groups 
and all 25 age combinations of the destination-source groups, separately 
for females and males.
Figure~\ref{fig:plotcorf} shows the resulting 
correlation matrix for the females as a heat map, left for time points before the 
lock-down, and right after lock-down.
The row and column labels are referring
to the group numbers.
For example, src1-3 refers to the time series
f15$\_$src -- f30$\_$dst, or dst5-1 is the series f75$\_$dst -- f15$\_$src.
The heatmaps show that the correlation structure before and after lock-down
has clearly changed.
Afterwards, there are more blocks with higher (absolute)
correlations, and thus more similarity or dissimilarity between certain
age groups.
In general, there is a more pronounced difference after lock-down
in the mobility behavior between the younger and the elder age groups.

\subsection{Incorporating spatial location}
\begin{figure}%[htbp]
\includegraphics[width=\linewidth]{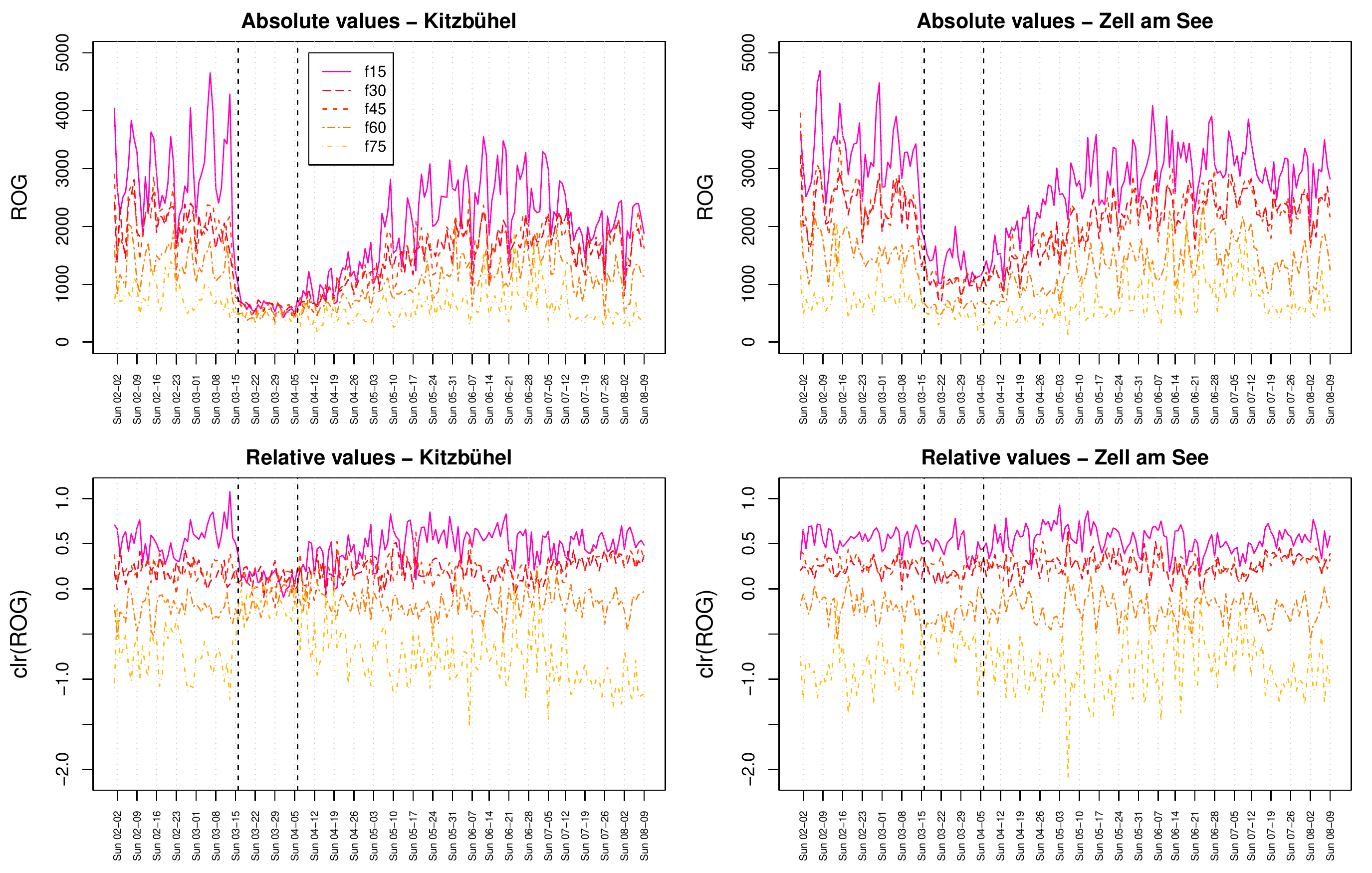}
\caption{Median \gls{rog} values for Kitzb\"uhel (left) and Zell am See (right)
as absolute (top) and relative (bottom) information.}
    \label{fig:plot_f50AB}
\end{figure}

The mobile phone data also contain information about the location, in
our case about the Austrian political district in which the phone has been used. 
The Austrian regions had different restrictions during the
lock-down phase, and in particular people from all districts in Tirol
had the strongest movement restrictions. 
Thus, in Figure~\ref{fig:plot_f50AB} we compare the median 
\gls{rog} values for Kitzb\"uhel, a district in Tirol, and Zell am See,
which is also a rural district but located in Salzburg. The absolute values
of the female age groups are shown in the upper plots, while the \gls{clr} coefficients are presented in the
lower plots.
Since the same scale is used along the vertical axes, one can clearly see the
difference in mobility during the lock-down period in Kitzb\"uhel and
Zell am See, and this is also visible in the \gls{clr} coefficients.
For Kitzb\"uhel, there is much smaller variability of the values 
during lock-down, and also the relative differences between the age groups 
become much smaller.
The change in the relative differences is not so
pronounced for Zell am See.
This means that also from a relative point of view,
the data structure changes completely in Kitzb\"uhel due to the restrictions.

\begin{figure}%[htbp]
\includegraphics[width=\linewidth]{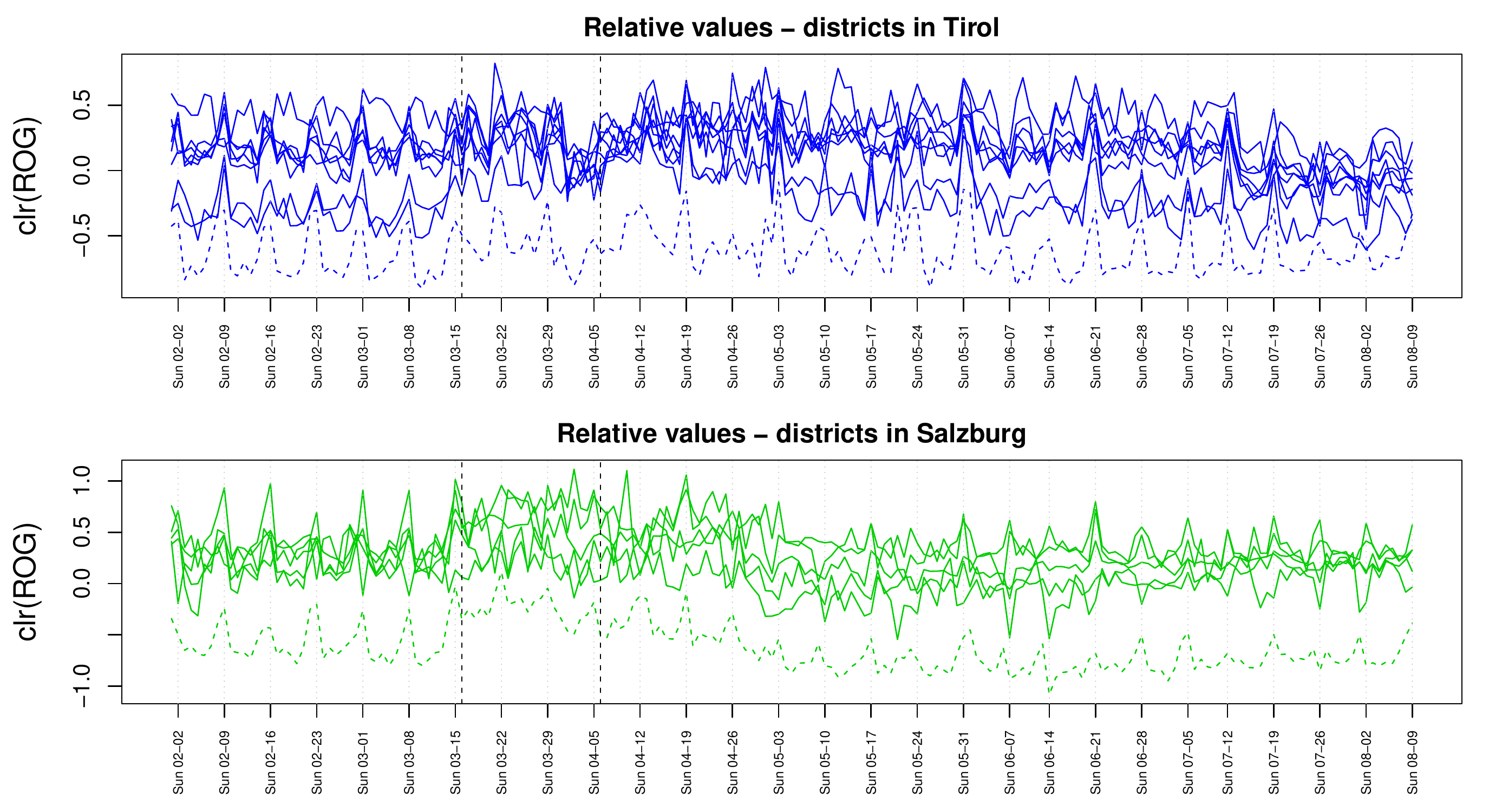}
\caption{\gls{clr} coefficients of median \gls{rog} values for the male group m30 in
all districts of Tirol
(top) and Salzburg (bottom).
The capitals are shown as dashed lines.}
    \label{fig:plotm30AB}
\end{figure}

Figure~\ref{fig:plotm30AB} focuses on the male age group m30, and compares the
composition of all districts in Tirol with that of all districts in Salzburg.
The dashed lines refer to the district capitals (Innsbruck and Salzburg, 
respectively).
These districts behave differently compared to the other districts
which are rural with many people commuting to their work place. 
The values of the districts in Tirol (except Innsbruck) get closer to each other
after lock-down, and they start to diverge only in the middle of April.
This may be explained by a similar mobility behavior of the m30 group within
this period, probably caused by home-office or reduced working time. This seems
different in districts of Salzburg, where the \gls{clr} coefficients show more
variability after lock-down.

\section{Discussion}
\label{sec:conclusions}

By analyzing relative changes using compositional data analysis methodologies formerly hidden insights can be identified.
In this work of analyzing mobility during the COVID-19 lock-down measures we see that certain age-groups of the population (elderly, young during weekends) do restrict mobility less than other members of the population.
Especially for the elderly which are at high risk of infections potentially additional reminders should be sent to adhere to the interventions.
Similarly, for the young groups on weekends additional reminders to use mouth nose protection could be useful.

%To be done.
%\ifdraftmode
%    {\color{red} TODO}
%\fi

\section*{Acknowledgements}
This work was funded by the WWTF under project COV COV20-035.

%Data is available upon request to the authors.

\section*{Declaration of interest}
We have no competing interests to declare.
%\printglossaries
\printbibliography

%\newpage
%\appendices
%Word count: 3880.
%\section{foo}
%\section{significance testing}

\end{document}